\documentclass[prl,aps,twocolumn,showpacs]{revtex4}
 \usepackage{epsfig}
 \usepackage{amssymb}
 \begin{document}
 \title{Quantum holography and multiparticle
 entanglement using ground-state coherences}
 \author{G. S. Agarwal$^{1}$,P. Lougovski$^{2}$,H. Walther$^{2}$}
 \affiliation{
     $^1$Physical Research Laboratory, Navrangpura,
     Ahmedabad, India\\
     $^2$Max-Planck-Institut f{\"u}r Quantenoptik, Hans-Kopfermann-Strasse
     1, D-85748 Garching, Germany }
 \begin{abstract}
 A method of storing and retrieving quantum states of radiation fields
  using the ground-state coherences is discussed. We
 demonstrate the generation of multiparticle entangled states
 starting from atoms prepared in a coherent state. Use is made of the j=-1/2 to
 +1/2 atomic transition and its interaction with a far-detuned
 elliptically polarized field.
 \end{abstract}
  \vspace*{-0.5cm}
 \maketitle
 
Storage and retrieval of light pulses has been extensively
 studied both theoretically and experimentally \cite{FLEISCHHAUER,LIU,ZIBROV,MATSKO,KUZMICH,DENG,VANDERWAL}. The
 light pulses can be stored in atomic coherences and later the
 atomic coherences can be converted back into the original light
 pulses \cite{FLEISCHHAUER,DENG}. The original work has been
 extended to storage of pulses of moderate
 powers \cite{TARAK,ARKHIPKIN}. A natural question arises: can one
 store and retrieve the quantum state of the field? This is the
 question addressed in this letter. It is shown how coherently
 prepared atomic systems interacting with far-detuned
 elliptical fields can serve the purpose. It is also demonstrated
 how the atomic coherences can be used for generating a
 variety of nonclassical states, including multiatom entangled
 states. Our model consists of N identical two-level atoms, with hyperfine structure splitting as depicted in Fig.1, coupled to
 two mutually orthogonal modes of a quantized electromagnetic field
 described by creation (annihilation) operators $a^\dagger_{-},
 a^\dagger_{+}$ ($a_{-}, a_{+}$) with the ground and excited-state coupling strengths $\Omega_{+}$ and $\Omega_{-}$, respectively. The Hamiltonian for the atom-field system reads
 \begin{eqnarray}
 \label{hamiltonian}H & = & \hbar\frac{\omega_{1}}{2}\sum^N_{j=1}
 (|e_{-}\rangle\langle e_{-}| - |g_{+}\rangle\langle g_{+}|)_{j} \nonumber \\
  & + & \hbar\frac{\omega_{2}}{2}\sum^N_{j=1}(|e_{+}\rangle\langle e_{+}|
   - |g_{-}\rangle\langle g_{-}|)_{j} \nonumber \\
  & + & \hbar\omega_{f}(a^{\dagger}_{+}a_{+} + a^{\dagger}_{-}a_{-}) \\
  & + & \hbar (\Omega_{-}a_{-}\sum^N_{j=1}|e_{-}\rangle\langle g_{+}|_{j} + h.c.) \nonumber \\
  & + & \hbar (\Omega_{+}a_{+}\sum^N_{j=1}|e_{+}\rangle\langle g_{-}|_{j} + h.c.) , \nonumber
 \end{eqnarray}
 where $\omega_{1}$ and $\omega_{2}$ are the frequencies of the transitions
 $|g_{+}\rangle\leftrightarrow|e_{-}\rangle$ and
 $|g_{-}\rangle\leftrightarrow|e_{+}\rangle$, respectively, and
 $\omega_{f}$ is the field frequency. In the dispersive limit of the
 atom-field coupling, i.e. when
 $\omega_{1} - \omega_{f} = \omega_{2} - \omega_{f} = \Delta\gg\{|\Omega_{+}|,|\Omega_{-}|\}$,
  one can reduce the Hamiltonian Eq.(\ref{hamiltonian}) to
 \begin{eqnarray}
 \label{effect}H_{1} & = &
 \hbar\frac{|\Omega_{-}|^{2}}{\Delta} a^{\dagger}_{-} a_{-}
 \sum^N_{j=1}(|e_{-}\rangle\langle e_{-}| - |g_{+}\rangle\langle g_{+}|)_{j} \nonumber \\
  & + & \hbar\frac{|\Omega_{+}|^{2}}{\Delta}a^{\dagger}_{+}a_{+}\sum^N_{j=1}(|e_{+}
  \rangle\langle e_{+}| - |g_{-}\rangle\langle g_{-}|)_{j} \\
  & + & \frac{2\hbar}{\Delta}(|\Omega_{-}|^2 \sum^N_{j=1}|e_{-}\rangle
  \langle e_{-}|_{j} + |\Omega_{+}|^2 \sum^N_{j=1}|e_{+}\rangle\langle e_{+}|_{j}). \nonumber
 \end{eqnarray}
 The effective Hamiltonian Eq.(\ref{effect}), in contrast
 to the original Eq.(\ref{hamiltonian}), does not couple
  ground states $\{|g_{-}\rangle_{j}, |g_{+}\rangle_{j}\},
   j=1,N$ to the excited ones. Therefore, if one starts
    from the initial atomic ground state, the only
    relevant contribution from the Hamiltonian Eq.(\ref{effect}) is
 \begin{eqnarray}
 \label{intermediate}H_{r} & = & -\hbar\frac{|\Omega_{-}|^{2}}{\Delta}
 a^{\dagger}_{-}a_{-}\sum^N_{j=1}(|g_{+}\rangle\langle g_{+}|)_{j} \nonumber \\
  & - & \hbar\frac{|\Omega_{+}|^{2}}{\Delta}a^{\dagger}_{+}a_{+}\sum^N_{j=1}
  (|g_{-}\rangle\langle g_{-}|)_{j}.
 \end{eqnarray}
 Introducing the collective atomic operator
 \begin{eqnarray}
 \label{colat}\hat{R_{z}} & = & \frac{1}{2}\sum^N_{j=1}(|g_{+}\rangle
 \langle g_{+}| - |g_{-}\rangle\langle g_{-}|)_{j},
 \end{eqnarray}
 one can express the operators $\sum\limits^N_{j=1}(|g_{+}\rangle\langle
  g_{+}|)_{j}$ and $\sum\limits^N_{j=1}(|g_{-}\rangle\langle g_{-}|)_{j}$
   in terms of $\hat{R_{z}}$ as follows:
 \begin{eqnarray}
 \frac{1}{2} \pm \hat{R_{z}} & = & \sum^N_{j=1}(|g_{\pm}\rangle\langle
 g_{\pm}|)_{j}. \nonumber
 \end{eqnarray}
 For the sake of simplicity hereafter we 
suppose that the two field modes have identical coupling strengths,
 i.e. $|\Omega_{+}| = |\Omega_{-}| = \Omega $.
 Finally, the effective Hamiltonian Eq.(\ref{effect}) will read
 \begin{eqnarray}
 \label{workh}H^{\prime} & = & \frac{\hbar \Omega^{2}}{\Delta}\hat{N}_{z}\hat{R}_{z},
 \end{eqnarray}
 where we have introduced the following field operator:
 \begin{eqnarray*}
 \hat{N}_{z} & = & a^{\dagger}_{+}a_{+} - a^{\dagger}_{-}a_{-} .
 \end{eqnarray*}
 Apart from the $\hat{R}_{z}$ operator, we construct the lowering $\hat{R}_{-}$ and raising $\hat{R}_{+}$ operators,
 \begin{eqnarray}
 \label{rmin}\hat{R}_{-} & = & \sum^N_{j=1}(|g_{-}\rangle\langle g_{+}|)_{j}, \\
 \label{rplus}\hat{R}_{+} & = & \sum^N_{j=1}(|g_{+}\rangle\langle g_{-}|)_{j}.
 \end{eqnarray}
 The operators $\hat{R}_{\pm}$ and $\hat{R}_{z}$ fulfill the following commutation relations:
 \begin{equation}
 \label{comrel}[\hat{R}_{-},\hat{R}_{+}] = -2\hat{R}_{z} ,
 [\hat{R}_{z},\hat{R}_{\pm}] = \pm\hat{R}_{\pm},
 \end{equation}
 and therefore describe an effective spin $J=\frac{N}{2}$ system,
 where $N$ is the number of atoms in the sample. As next step
 we define the eigenvectors of the z component of the total spin
 operator $\hat{R}_{z}$ as follows:
 \begin{equation}
 \label{eigenvect}\hat{R}_{z}|m\rangle = \hbar m|m\rangle, J=\frac{N}{2},
  m =-J,-J+1,\cdots,J-1,J.
 \end{equation}
 Action of the lowering (raising) operators $\hat{R}_{-}$ ($\hat{R}_{+}$)
  on the eigenvectors of $\hat{R}_{z}$ can be calculated in a
  straightforward manner:
 \begin{eqnarray}
 \hat{R}_{-}|m\rangle & = & \hbar\sqrt{(J+m)(J-m+1)}|m-1\rangle, \\
 \hat{R}_{+}|m\rangle & = & \hbar\sqrt{(J-m)(J+m+1)}|m+1\rangle.
 \end{eqnarray}
 The set of eigenvectors $|m\rangle, m=\overline{-J,J}$ of the operator
 $\hat{R}_{z}$ is trivial connected to the previously introduced set
  $\{|g_{-}\rangle_{j}, |g_{+}\rangle_{j}\}, j=\overline{1,N}$,
  which denotes the hyperfine structure of the ground states of atoms.
   For example $|-J\rangle = \prod^{N}_{k=1}|g_{-}\rangle_{k}$, and
   consequently we will call the state $|-J\rangle$ the ground state.
 It is instructive to introduce the following field operators,
 \begin{equation}
 \hat{N}_{-} = a^{\dagger}_{-}a_{+}, \hat{N}_{+} = a^{\dagger}_{+}a_{-},
 \end{equation}
 and compare the algebraic properties of the two sets of operators,
 ${\hat{N}_{z}, \hat{N}_{-}, \hat{N}_{+}}$ and ${\hat{R}_{z},
 \hat{R}_{-}, \hat{R}_{+}}$. Operators $\hat{N}_{z}$,
 $\hat{N}_{-}$ and $\hat{N}_{+}$ satisfy the same commutation
  relations as in Eq.(~\ref{comrel}), i.e.
 \begin{equation}
 \label{comrel1}[\hat{N}_{-},\hat{N}_{+}] = -2\hat{N}_{z} ,
 [\hat{N}_{z},\hat{N}_{\pm}] = \pm\hat{N}_{\pm}.
 \end{equation}
 Therefore, coming back to the Hamiltonian Eq.(\ref{workh}), one can say
 that there is effectively a spin-spin interaction along the z-axis, where
 one of the spins is the collective spin of the atomic sample and the other
 one is some effective spin which corresponds to two orthogonally polarized
 modes of the cavity field.
 As a result, the Hamiltonian Eq.(\ref{workh}) is very suitable for 
spin quantum nondemolition measurement (QND) as well as being a good candidate
 for the generation of spin-squeezed atomic samples~\cite{Polzik}.
 \begin{figure}[!ht]
 \begin{center}
 \includegraphics[width=68mm]{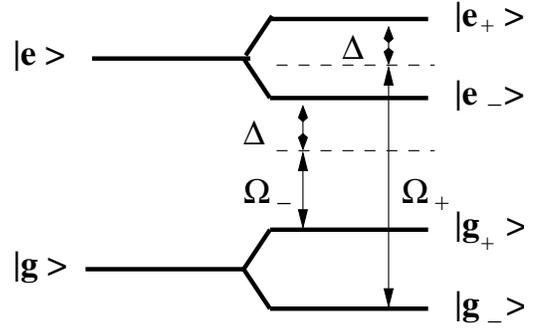}
 \caption{\label{atstruct} Internal atomic stucture }
 \end{center}
 \end{figure}
 
We now demonstrate important applications of the Hamiltonian
 Eq.(\ref{workh}) to create mesoscopic superposition states
 ~\cite{Schrodinger}, multiatom entangled states and to store
 and retrieve quantum states.
 
{\bf Mesoscopic superposition of states:} Let us consider the case where the atomic sample has been initially prepared in some coherent atomic 
state~\cite{Arecchi}
 \begin{eqnarray}
 \label{atstate} && |\theta,\phi\rangle = R_{\theta,\phi}|-J\rangle \\
 & = & \hspace*{-0.3cm}\sum^J_{m=-J}{2J\choose
 J+m}^{\frac{1}{2}}\left[\cos\left(\frac{\theta}{2}\right)\right]
 ^{J-m}\left[\sin\left(\frac{\theta}{2}\right)e^{-i\phi}\right]^{J+m}\hspace*{-.5cm}|m\rangle.
 \nonumber
 \end{eqnarray}
 Here $R_{\theta,\phi}$ is the rotation operator which rotates the
  initial ground state $|-J\rangle$ by an angle $\theta$ around an
  axis $\hat{m}=(\sin(\phi),-\cos(\phi),0)$. From the practical
  point of view the coherent atomic state in Eq.(\ref{atstate}) can be prepared
 in a number of ways; for example, by applying a homogeneous
 magnetic field along the $\hat{m}$ axis to the ground state of the
 atomic sample $|-J\rangle$. The other interesting and popular
 possibility is to use short pulses and Raman transitions, a
 technique that has been extensively used by, for example, Wineland and
 co-workers \cite{WINELAND}. We consider interaction of the coherently
 prepared atomic sample with an elliptically polarized field which
 we represent by the state $|\psi_{f}\rangle =
 \alpha|1_{+},0_{-}\rangle + \beta|0_{+},1_{-}\rangle$,
 ($|\alpha|^{2}+|\beta|^{2} = 1$). Note that $\alpha=0$ or
 $\beta=0$ corresponds to a circularly polarized field. A
 calculation by means of the interaction Hamiltonian Eq.(\ref{workh}) leads
 to the following atom-field state
 \begin{eqnarray}
 |\Psi_{at-f}\rangle & = & \exp(-i\phi_{0}t\hat{N}_{z}\hat{R}_{z})
 |\theta,\phi\rangle|\psi_{f}\rangle \\
 \label{cat} & = &
 \alpha|\theta,\phi+\phi_{0}t\rangle|1_{+},0_{-}\rangle
 e^{i\phi_0tJ} \nonumber\\&+& \beta|
 \theta,\phi-\phi_{0}t\rangle|0_{+},1_{-}\rangle e^{-i\phi_0tJ}.
 \nonumber
 \end{eqnarray}
 Here $\phi_{0} = \Omega^{2}/\Delta$ and $t$ is the interaction time. The
 atom-field state Eq.(\ref{cat}) describes a situation when atoms
 and the cavity field get entangled. Detection of the field
 polarization in the direction of, say, $x$ leads to an atomic state
 \begin{eqnarray}
 |\Psi\rangle_{cat} &=& \alpha
 e^{i\phi_0tJ}|\theta,\phi+\phi_{0}t\rangle\nonumber\\
  &+& \beta e^{-i\phi_0tJ}|\theta,\phi-\phi_{0}t \rangle .
 \end{eqnarray}
 This corresponds to a mesoscopic superposition
 of coherent states for the atomic system. This is the case as long as 
the two coherent states in the above equation are not orthogonal, i.e.
 \begin{eqnarray}
 \label{overlap}\langle\theta,\phi+\phi_{0}t|\theta,
 \phi-\phi_{0}t \rangle & = & \exp(2iJ\phi_{0}t)
 [\cos(\phi_{0}t) \nonumber\\
 & - & i\cos(\theta)\sin(\phi_{0}t)]^{2J}.
 \end{eqnarray}
 is not zero. From Eq.(\ref{overlap}) follows
 that for a particular choice of parameters
 (for example, $\phi_{0}t=\frac{\pi}{2}$ and $\theta=\frac{\pi}{2}$,
  $\forall \phi$) the overlap between coherent atomic states
  is zero and hence, the atom-field state in Eq.(\ref{cat})
  becomes a mixed state. We have thus shown how the dispersive
  interactions on ground-state atoms can produce long-lived
  mesoscopic superpositions.
 We note that dispersive interaction has
  been considered in the context of two-level atoms to produce
  both atomic and field cat states \cite{RP,HAROCHE,ZENG,GERRY}.
 
{\bf Generation of a multiatom entangled state:} It is next shown that the
 above mesoscopic superposition is equivalent to a multiatom
 entangled state. For this purpose we express the coherent state of
  Eq. $(14)$ as a product state over all the atoms:
 \begin{eqnarray}
 &&|\theta,\phi\rangle=\Pi_{j=1,N}|\theta,\phi\rangle_{j}\nonumber\\
 &&|\theta,\phi\rangle_{j}=\sin(\theta/2)e^{-i\phi}|g_{+}\rangle_{j}
 +\cos(\theta/2)|g_{-}\rangle_{j},
 \end{eqnarray}
 and hence the mesoscopic superposition $(16)$ can be expressed as
 \begin{eqnarray}
 |\psi\rangle_{cat}&=&\alpha\Pi_{j=1,N}|\theta,\phi+\phi_{0}t\rangle_{j}
 e^{i\phi_{0}tJ}\nonumber\\
 &+&\beta e^{-i\phi_{0}tJ}\Pi_{j=1,N}
 |\theta,\phi-\phi_{0}t\rangle_{j}.
 \label{fprob}
 \end{eqnarray}
 This is a multi-atom entangled state. We further note that
 $\langle\theta,\phi-\phi_{0}t|\theta,\phi+\phi_{0}t\rangle=0$ if
 $\phi_{0}t=\pi/2,~\theta=\pi/2$. Thus the two parts of $(19)$
 become orthogonal for this choice of parameters. In this case
 $(19)$ becomes an example of GHZ states~\cite{GHZstate}. It has thus been shown how long-lived GHZ states can be produced by using ground-state atomic
 coherences.
 
{\bf Interaction-free evolution of field states:}
  It is clear from Eq.$(15)$ that the probability of
  detecting the field state unchanged after the coupling to atoms
  strongly depends on the atom-field interaction time. If we
  choose $\phi_{0}t=\pi$ and make the number of atoms in the sample N
   even, the cavity field after interaction is in the
  same state as the initial cavity field. This is reminiscent of the trapping states
  in a micromaser \cite{TRAPPING} where for certain field states the atoms leave the cavity in the original states. To verify the above statement we consider the evolution
  operator
  $exp\{-i\pi \hat{N_{z}}\hat{R_{z}}\}$ for $\phi_{0}t=\pi$. Since
  $\hat{N_{z}}$ has the eigenvalues $\pm 1$ and for even $N$, $\hat{R_{z}}$ has an integer
  eigenvalue and the above evolution operator is the same for the two eigenvalues of
  $\hat{N_{z}}$. Thus the field would revert to the original state. For odd $N$,
  the eigenvalues of $\left(\frac{N}{2}+\hat{R_{z}}\right)$ are
  integers and then the two evolution operators $exp\{\mp
  i\pi\hat{R_{z}}\}$ will differ by an overall phase factor
  $\pi~~\left(exp\{2i\pi N/2\}\right)$. In this case the state of
  the output field will be
  $\alpha|1_{+},0_{-}\rangle-\beta|0_{+},1_{-}\rangle$.
 
{\bf Quantum holography :} It is next demonstrated how our system can
 be used to store the quantum state of the field via atomic
 coherence . This is the quantum counterpart of holography in which
 the object information is stored in a hologram using a reference
 field. The hologram when irradiated by a coherent field,
 recovers the information on the object. Clearly Eq.$(15)$
 already shows how the information about the field state is stored
 in atomic ground-state coherence. Note that this is long-lived
 storage since we use the ground states of the atomic systems. It is now 
shown how this quantum hologram can be read to retrieve the
 information on the field. One possibility is to measure the atomic
 population in the state $|g_{-}\rangle$. This is easily done by
 applying a laser pulse polarized appropriately so as to move the
 population from $|g_{-}\rangle$ (and not from $|g_{+}\rangle$) to
 an excited state and to measure the subsequent fluorescence. This measurement
 will project the state $(15)$ on to a state of the field given by
 \begin{eqnarray}
 |\psi_{f}\rangle_{-J}&&\rightarrow
 \left(\cos{\frac{\theta}{2}}\right)^{2J}\times\nonumber\\
 &&\left(\alpha|1_{+},0_{-}\rangle
 e^{i\phi_{0}tJ}+\beta|0_{+},1_{-}\rangle e^{-i\phi_{0}tJ}\right).
 \end{eqnarray}
 Thus the stored state of the field is recovered apart from a
 relative phase factor $2J\phi_{0}t$. Note that this phase factor
 is known a priori from the preparation process of the hologram.
 
 {\bf Generation of cat states for the radiation field:} Finally, it is shown how the
  interaction $(5)$ is quite suited to production of cat
  states for the radiation field. This possibility has already been
  extensively discussed by Haroche and co-workers\cite{HAROCHE} using Rydberg
  atoms. Here we use ground-state coherence. Consider, for example,
  a single atom $J=1/2$ interacting with a field in coherent state
  $|\alpha,\beta\rangle$. Obviously, the state of the combined system
  at time $t$ will be
  \begin{eqnarray}
  |\psi(t)\rangle&\equiv & exp\{-i\phi_{0}t\hat{N_{z}}\hat{R_{z}}\}
  |\alpha,\beta\rangle|\theta,\phi\rangle\nonumber\\
  &=&\cos\frac{\theta}{2}|g_{-}\rangle|\alpha e^{i\phi{_0}t/2},\beta
  e^{-i\phi{_0}t/2}\rangle\nonumber\\
  &+& e^{-i\phi}\sin\frac{\theta}{2}|g_{+}\rangle|
  \alpha e^{-i\phi{_0}t/2},\beta e^{i\phi{_0}t/2}\rangle.
  \end{eqnarray}
  Clearly, measurement of the atomic population in a state
  different from $|g_{\pm}\rangle$ produces a cat state of the
  radiation field.
 
 In conclusion, it has been shown how ground-state coherence can be
  utilized for quantum holography and for generation of long-lived
  mesoscopic superpositions and multi particle GHZ states. The
  latter states possess very interesting nonclassical properties.

\end{document}